\documentclass[a4paper,11pt]{article}

\usepackage{pos}
\usepackage[capitalise]{cleveref}
\usepackage{slashed}
\usepackage[utf8]{inputenc}
\usepackage{graphicx}
\usepackage{subcaption}
\usepackage{hyperref}

\graphicspath{{./figures/}}

\newcommand{\Nf}{N_{\mathrm{f}}}
\newcommand{\ii}{\mathrm{i}}
\newcommand{\cc}{\langle\bar{\psi}\psi\rangle}
\newcommand{\muc}{\mu_c}
\newcommand{\psibar}{\bar{\psi}}
\newcommand{\sigex}{\langle\sigma\rangle}
\newcommand{\sigmabar}{\bar{\sigma}}
\newcommand{\Veff}{V_\mathrm{eff}}
\renewcommand{\L}{\mathcal{L}}

\title{The magnetized Gross-Neveu model at finite chemical potential}
\ShortTitle{Magnetized GN model at finite $\mu$}

\author*[a,b]{Michael Mandl}
\author[a,c]{Julian J. Lenz}

\affiliation[a]{Theoretisch-Physikalisches Institut, Friedrich-Schiller-Universität Jena,\\
Fröbelstieg 1, D-07743 Jena, Germany}
\affiliation[b]{Institute of Physics, NAWI Graz, University of Graz,\\ Universitätsplatz 5, A-8010 Graz, Austria}
\affiliation[c]{Swansea Academy of Advanced Computing, Swansea University,\\
Fabian Way, SA1 8EN, Swansea, Wales, UK}

\emailAdd{j.j.lenz@swansea.ac.uk}
\emailAdd{michael.mandl@uni-graz.at}

\abstract{
We study the $(2+1)$-dimensional Gross-Neveu model at non-zero chemical potential and subjected to a
homogeneous background magnetic field. We do so both analytically, in the limit of an infinite number
of fermion flavors in which mean-field approaches become exact, as well as on the lattice for a single 
flavor. The rich and exotic phase structure observed in the mean-field limit is found to be destroyed when strong 
quantum fluctuations are present in the system. Instead, in the phase of spontaneously broken chiral 
symmetry the magnetic field enhances this breaking for all choices of
parameters. As a byproduct, we find indications for a first-order phase
transition in the chemical potential for vanishing magnetic field but also
provide hints that this could rather be a finite-size than a finite-flavor-number
effect.}

\FullConference{The 40th International Symposium on Lattice Field Theory (Lattice 2023)\\
July 31st - August 4th, 2023\\
Fermi National Accelerator Laboratory\\}

\begin{document}
\maketitle

\section{Introduction}
	The study of strongly-interacting matter under extreme conditions has widespread applications, ranging
from the description of heavy-ion collisions or compact stellar objects to models of the early universe. 
In this context, `extreme conditions' refers to exceedingly high temperatures or densities, strong 
magnetic fields and combinations thereof. The strong interactions are described by the theory of Quantum 
Chromodynamics (QCD), whose behavior at zero and non-zero temperature and within background magnetic 
fields is well understood by now, owing to extensive numerical studies in the framework of lattice 
quantum field theory \cite{BBE12_2}. However, little is known about the finite-density regime of QCD due 
to a strong complex-action problem preventing the straightforward application of conventional lattice 
methods based on importance sampling. 

In order to nonetheless gain insight into the behavior of strongly-interacting matter at finite
density, one commonly reverts to the study of effective models expected to reproduce QCD phenomenology 
within their range of validity. One particular class of such model theories is constituted by four-Fermi 
theories, i.e., models of relativistic fermions with local four-point self-interactions. Apart from 
various applications in condensed-matter physics (see, e.g., \cite{CM94}), they have contributed 
substantially to our understanding of strongly interacting matter at finite temperature and density 
\cite{Kle92r}. Moreover, they have been shown to reproduce the inverse magnetic catalysis phenomenon 
observed in lattice simulations of QCD within background magnetic fields \cite{BBE12_2}, provided that 
the four-Fermi coupling runs appropriately with the magnetic field \cite{EM19}. Most model approaches, 
however, employ the mean-field approximation, which, despite its success, is not guaranteed to describe 
real physical systems faithfully due to the suppression of quantum fluctuations. In order to assess the 
effect of said fluctuations, one may, e.g., perform lattice simulations beyond the mean-field limit and 
this is, in fact, the approach we pursue in this work. Many comparable previous studies found that 
mean-field approaches generally provide a solid understanding of the qualitative features of a theory, 
with corrections arising only on a quantitative level when going beyond \cite{LPW20,LMW22,LMW23}. In 
this work, however, we present a counter-example to this assertion.
\section{The Gross-Neveu model}
	We study the simplest four-Fermi theory, the so-called Gross-Neveu model \cite{GN74}, 
\begin{equation}\label{eq:gn_lagrangian}
	\L = \psibar(x)\left(\slashed{\partial} + 
					     \sigma(x) + 
					     \mu\gamma_0 + 
					     \ii e\slashed{A}(x)\right)\psi(x) + 
		 \frac{\Nf}{2g^2}\sigma^2(x)\;,
\end{equation}
where the auxiliary scalar field $\sigma(x)$ was introduced in exchange for the scalar-scalar 
interaction term $\left(\psibar\psi\right)^2$ via a Hubbard-Stratonovich transformation as usual. In 
\eqref{eq:gn_lagrangian}, $\psi(x)$ is used to denote $\Nf$ flavors of massless fermion fields, $\mu$
is the fermion number chemical potential, $e$ denotes the elementary electric charge, $A_\mu$ controls 
an external magnetic field and $g^2$ denotes the four-Fermi coupling. The Lagrangian 
\eqref{eq:gn_lagrangian} is invariant under discrete chiral transformations of the form
\begin{equation}
	\psi(x) \to \ii\gamma_5\psi(x)\;, \quad 
	\psibar(x) \to \ii\psibar(x)\gamma_5\;, \quad
	\sigma(x) \to -\sigma(x)\;,
\end{equation}
and this symmetry is broken by a fermionic mass term. A breaking pattern like
this entails that one may employ the chiral condensate $\cc$,
which is related to the expectation value of $\sigma$ via 
\begin{equation}
	\cc = \frac{\ii\Nf}{g^2}\sigex\;,
\end{equation}
as an order parameter for chiral symmetry breaking.
Throughout, we shall work in $2+1$ (Euclidean) space-time dimensions and with four-component spinors. 
The magnetic field $B=F_{12}:=\partial_1A_2(x)-\partial_2A_1(x)$ is chosen to be constant and 
homogeneous. 

\section{The large-\texorpdfstring{$\Nf$}{Nf} limit}
	A first approach to studying the Gross-Neveu and similar models is to let the number of fermionic 
flavors tend to infinity, $\Nf\to\infty$. In fact, in this limit the mean-field approximation becomes 
exact and the problem of determining the phase structure in $(B,T,\mu)$ space is reduced to a 
minimization problem by means of a saddle point expansion terminated at the lowest order. More 
concretely, if we assume the scalar field to be independent of space and time, 
$\sigma(x)=\sigma=const.$, then $\sigex$ is given by the minimum position of the effective potential
\begin{equation}
	\Veff(\sigma) = \frac{\sigma^2}{2g^2} - 
					\frac{1}{V}\ln\det\left(\slashed{\partial} + 
									     	\sigma + 
								 			\mu\gamma_0 + 
								 			\ii e \slashed{A}\right)\;,
\end{equation}
where $V=\beta L^2$ denotes the space-time volume, $\beta=1/T$ is the inverse temperature and $L$ 
denotes the extent of space in each direction. 
\begin{figure}
	\centering
	\includegraphics[width=0.9\textwidth]{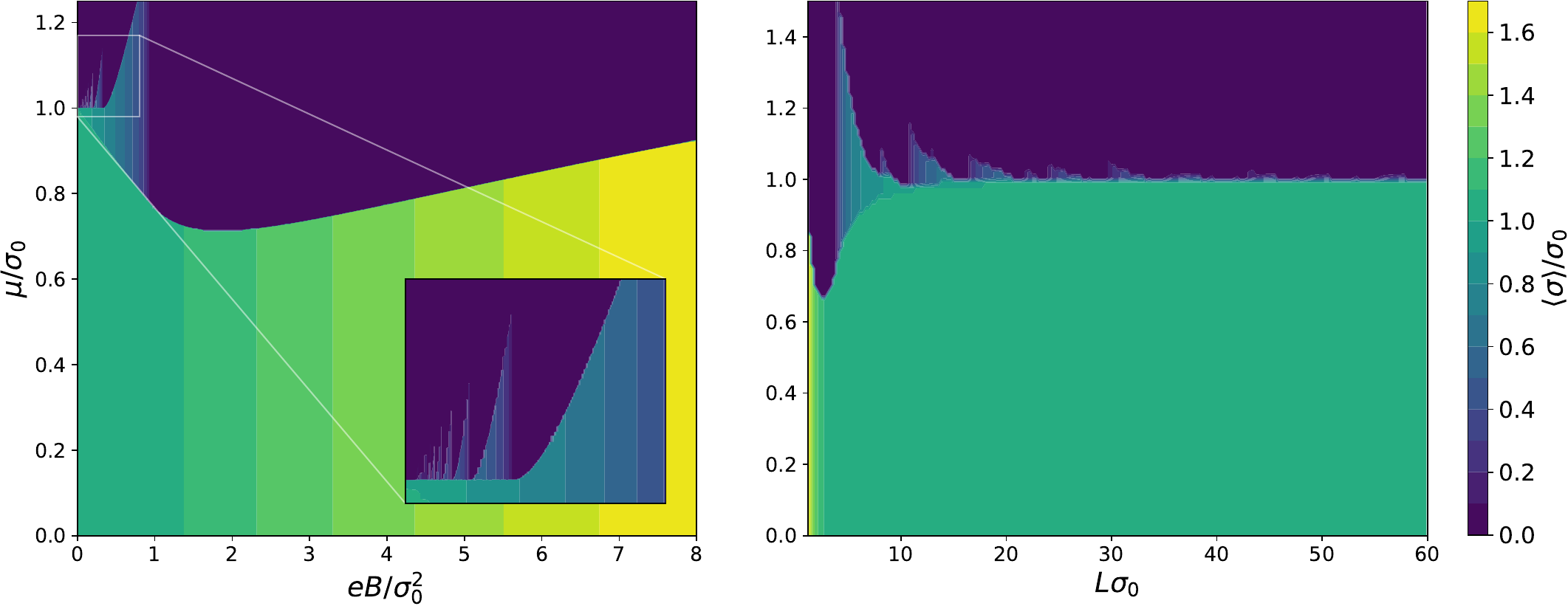}
	\caption{Phase diagrams in the mean-field limit at zero temperature. Left: $L=\infty$. Right: 
			 $B=0$. The scale $\sigma_0$ is set by the value of $\sigex$ at vanishing $B$, $T$, and 
			 $\mu$ and in an infinite volume.}
	\label{fig:pd_large_N}
\end{figure}

This effective potential can be computed in closed form, see, e.g., \cite{LMW23_2} and its minimization  
allows for the study of the model's phase structure. The situation at vanishing chemical potential was 
treated exhaustively in \cite{LMW23}, whereas we are concerned with non-zero $\mu$ but zero temperature 
here. In the following, $g^2$ is assumed to be larger than a critical value, in such a way that chiral 
symmetry is spontaneously broken at vanishing $B$, $T$, and $\mu$. We show the phase diagram of 
the model in $(B,\mu)$-space in the infinite-volume limit in \cref{fig:pd_large_N} (left). One observes 
that for low $\mu$ the magnetic field enhances chiral symmetry breaking, a phenomenon referred to as 
magnetic catalysis \cite{GMS95}. At larger $\mu$, however, a small region emerges in which the magnetic
field instead has the opposite effect and one finds so-called inverse magnetic catalysis \cite{PRS11}. 
Perhaps most interestingly, close to the region of inverse magnetic catalysis, the formation of Landau 
levels in the model gives rise to a pattern of multiple (first-order) phase transitions in $\mu$ 
between the phase of broken chiral symmetry and the symmetric phase. It is interesting to note that a 
non-zero magnetic field can induce first-order transitions which are not present at $B=0$ in the 
infinite-volume limit.

For $L<\infty$, on the other hand, the discretization of spatial momenta also gives rise to a fully 
discrete energy spectrum, in close analogy to the Landau quantization. This analogy can be appreciated 
from \cref{fig:pd_large_N} (right), where we plot the phase diagram in the $(L,\mu)$ plane for $B=0$. 
Similar to the situation at non-zero magnetic field, one thus observes multiple phase transitions which 
can be of first order and the same is also found for low non-zero temperatures \cite{LMW23_2}. Notice
that, despite the finite spatial volume, these transitions are still proper phase transitions since
we work in the large-$\Nf$ limit and $\Nf$ and $V$ enter the path integral in an analogous way. In
what follows, we study what happens to the mean-field phase structure presented in 
this section when considering a finite number of fermion flavors, i.e., going beyond the mean-field 
limit.

\section{Lattice results}
	\begin{figure}
	\centering
	\includegraphics[width=0.8\textwidth]{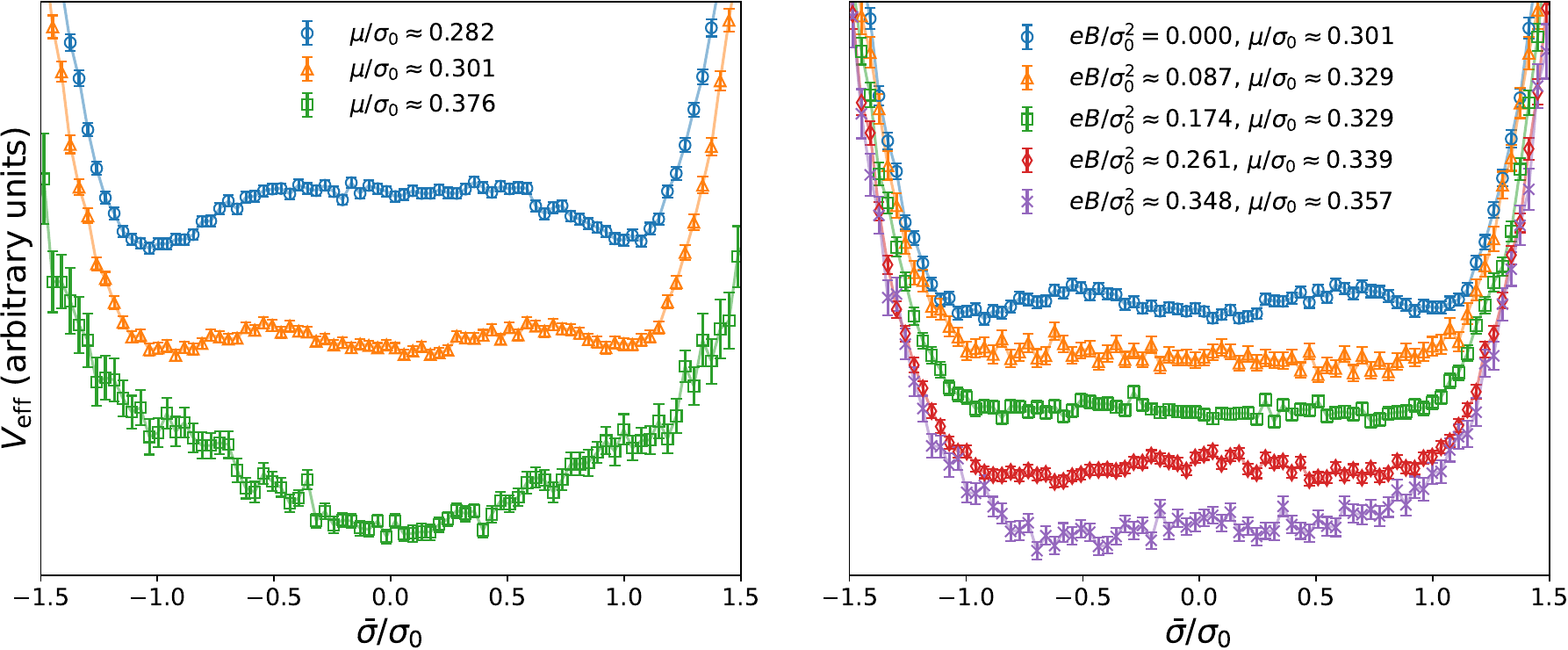}
	\caption{Constraint effective potential on an $8^3$ lattice at $T/\sigma_0\approx0.118$ with   	
			 $a\sigma_0\approx1.063$ ($a$ denotes the lattice spacing). Left: $B=0$. Right: $B\neq0$ 
			 and $\mu$ close to the phase transition. The potentials are plotted in arbitrary units and 
			 shifted vertically for visual clarity.}
	\label{fig:veff_lattice}
\end{figure}

In order to address this question, we study the model on the lattice; in particular, we perform 
simulations, employing Neuberger's overlap Dirac operator \cite{Neu98}, at $\Nf=1$ in order to deviate 
from the large-$\Nf$ limit as much as possible. For details on our simulation setup, measured 
observables and scale-setting, as well as for a list of all parameter values we have performed 
simulations for, we refer to \cite{LMW23_2}. This reference also outlines how the complex-action 
problem arising in our simulations is avoided.

We first consider the case of vanishing magnetic field but non-zero $\mu$, for which previous lattice 
simulations employing staggered fermions have conjectured the existence of a first-order phase 
transition at low non-zero temperatures \cite{KS01}. This observation was later explained to be a 
consequence of going beyond the mean-field limit \cite{KPR07}. However, given our remarks in the 
previous section and the fact that simulations can be performed only in finite volumes, it is not 
entirely clear what the true origin of this first-order phase transition is, since a similar phenomenon 
can also be observed in the large-$\Nf$ limit on a finite volume. After all, one
typically expects quantum 
fluctuations to weaken phase transitions rather than strengthen them.

In order to study the possible existence of a first-order transition in the model at $B=0$ we show in 
\cref{fig:veff_lattice} (left) the (appropriately normalized) logarithm of the probability distribution 
of $\sigmabar:=\frac{1}{V}\sum_x\sigma(x)$, corresponding to the (constraint) effective potential, for 
different values of $\mu$. As one can see, the effective potential has two degenerate minima at low 
$\mu$ but develops a third one around $\mu/\sigma_0\approx0.301$, while for large enough $\mu$ it has 
only a single minimum at zero. This behavior is reminiscent of the mean-field limit and could hint at a 
first-order transition. However, since both $\Nf$ and $V$ are small in our simulations, this finding is 
not fully conclusive. The situation is somewhat different when considering non-zero magnetic fields, as 
seen in \cref{fig:veff_lattice} (right), where we plot the effective potential for various values of 
$B$ and for $\mu$ close to the phase transition. We no longer see clear evidence for a third minimum in
the potential, which rather suggests a second-order phase transition in the thermodynamic limit. 

\begin{figure}
	\centering
	\includegraphics[width=0.98\textwidth]{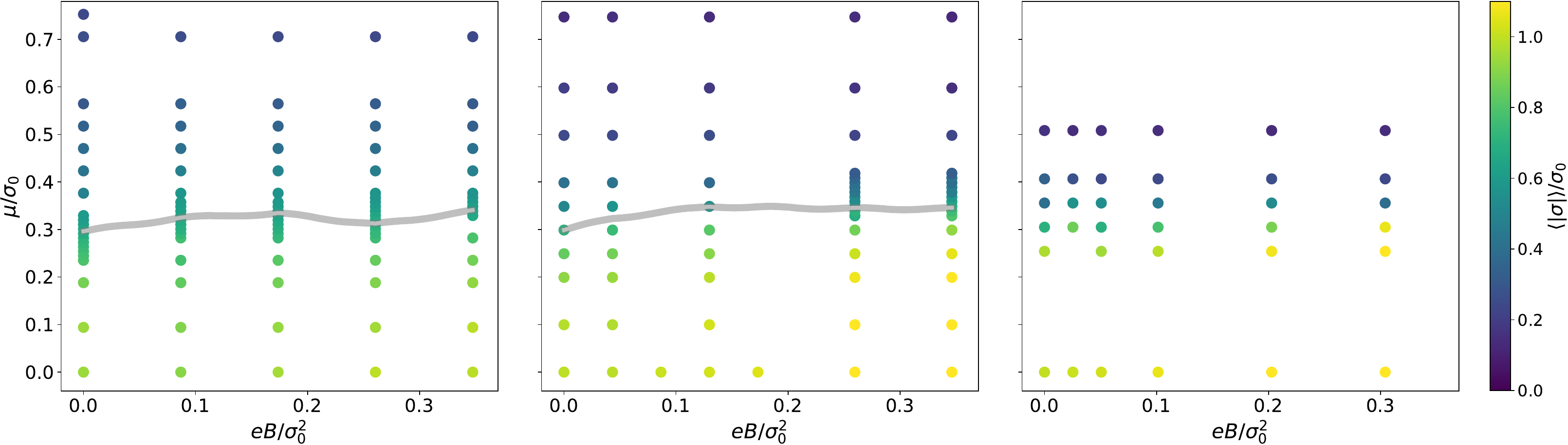}
	\caption{$(B,\mu)$ phase diagram for increasing physical volume and fixed lattice spacing. Left: 
	$8^3$ lattice, $a\sigma_0\approx1.063$. Center: $12^3$ lattice, $a\sigma_0\approx1.004$. Right: 
	$16^3$ lattice, $a\sigma_0\approx0.984$. The thick lines show a rough estimate of the critical 
	chemical potential.}
	\label{fig:pd_infVol}
\end{figure}

\begin{figure}
	\centering
	\includegraphics[width=0.94\textwidth]{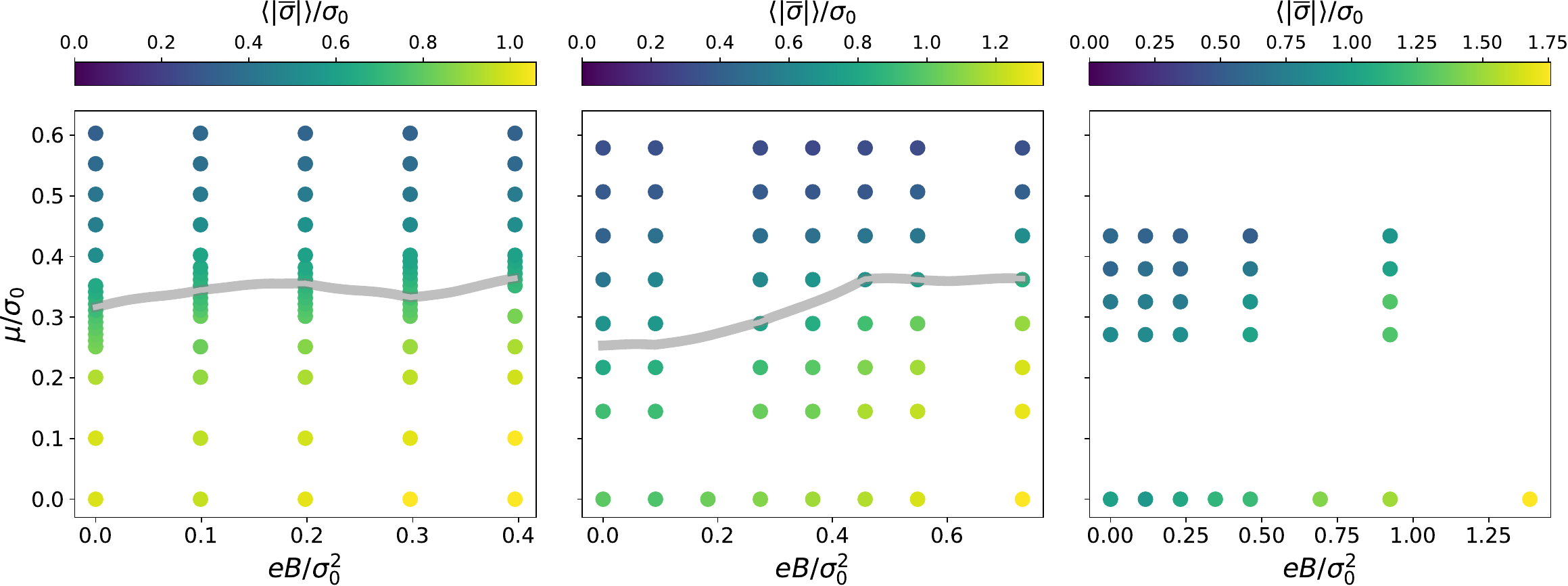}
	\caption{$(B,\mu)$ phase diagram for decreasing lattice spacing and fixed physical volume. Left: 
	$8^3$ lattice, $a\sigma_0\approx0.995$. Center: $12^3$ lattice, $a\sigma_0\approx0.691$. Right: 
	$16^3$ lattice, $a\sigma_0\approx0.451$. The thick lines show a rough estimate of the critical 
	chemical potential.}
	\label{fig:pd_cont}
\end{figure}

Let us now study the phase structure of the model in $(B,\mu)$ space. The analog of 
\cref{fig:pd_large_N} (left) for $\Nf=1$ is presented in \cref{fig:pd_infVol,fig:pd_cont}; the former 
shows increasing volumes at fixed lattice spacing $a$, while the approach to the continuum limit at 
constant volume is depicted in the latter. For the smaller lattices, where the data allowed for a rough 
estimate of the critical chemical potential $\muc$ of the transition, we have indicated the dependence 
of this estimate on $B$ as thick gray lines. One observes that, apart from finite-size effects the 
magnetic field enhances chiral symmetry breaking for all values of $\mu$ below the transition and the 
critical chemical potential increases with $B$. Thus, our data are consistent with magnetic catalysis 
for all $\mu<\muc$. Moreover, we do not observe any trace of multiple phase transitions in $\mu$ for 
finite $B$. These findings are in stark contrast to the large-$\Nf$ expectations and also contradict 
the analytical study \cite{KPR13} working at finite $\Nf$. We believe that this discrepancy arises due 
to strong fluctuations of $\sigma(x)$ present at $\Nf=1$ but absent in the mean-field limit and in 
\citep{KPR13}. Further investigations along this line of research are currently
underway. It could, however, also be possible that our current approach simply
does not allow for a sufficiently fine parameter scan with sufficiently high
statistics to resolve these delicate features. Estimates on this are presented
in \cite{LMW23_2}.

\section{Summary \& Outlook}
	We have investigated the Gross-Neveu model \eqref{eq:gn_lagrangian} in $2+1$ dimensions, both 
analytically in the limit of a large number of fermion flavors, as well as on the lattice using $\Nf=1$ 
flavor of overlap fermions. In the mean-field limit, we find a rich phase structure in the parameter 
space spanned by the temperature, the chemical potential, a homogeneous background magnetic field, and 
the spatial volume. In particular, at non-zero magnetic field this phase structure features both 
magnetic catalysis and inverse magnetic catalysis, as well as multiple phase transition patterns. 
Our lattice simulations, on the other hand, indicate that for $\Nf=1$ the situation is drastically 
different and, in fact, much simpler: We find only magnetic catalysis to be present for all chemical 
potentials below some critical value, the latter increasing with $B$, and no evidence for multiple 
transitions. This investigation thus provides a counter-example for the common lore that the large-$\Nf$ 
limit usually serves as a reliable guiding principle for the study of the qualitative behavior of a 
four-Fermi theory. It would be interesting to test this observation also in more realistic models of 
QCD, which up to now have been studied mostly from the mean-field point of view.

\acknowledgments
M. M. thanks Gergely Endr\H{o}di and Laurin Pannullo for enlightening conversations. The authors are 
indebted to Andreas Wipf for multiple useful discussions and to Björn Wellegehausen for providing the 
simulation codebase underlying these studies. This work has been funded by the Deutsche 
Forschungsgemeinschaft (DFG) under Grant No. 406116891 within the Research Training Group RTG 2522/1. 
The work of J. J. L. was supported by the UKRI Science and Technology Facilities Council (STFC) 
Research Software Engineering Fellowship EP/V052489/1 and by the Supercomputing Wales project, which is 
part-funded by the European Regional Development Fund (ERDF) via Welsh Government. The simulations were 
performed on resources of the Friedrich Schiller University in Jena supported in part by the DFG Grants 
No. INST 275/334-1 FUGG and No. INST 275/363-1 FUGG, as well as on the Swansea University SUNBIRD 
cluster (part of the Supercomputing Wales project). The Swansea University SUNBIRD system is part 
funded by the European Regional Development Fund (ERDF) via Welsh Government.
    
\bibliographystyle{JHEP}
\bibliography{bibliography}

\end{document}